\documentclass{sigchi}

\setcopyright{acmcopyright}

\doi{http://dx.doi.org/10.1145/3051457.3054013}
\isbn{978-1-4503-4450-0/17/04}
\conferenceinfo{LAS'17,}{April 2017, Boston, MA, USA}
\CopyrightYear{2017}
\acmPrice{\$15.00}

\usepackage{balance}       
\usepackage{graphics}      
\usepackage[T1]{fontenc}   
\usepackage{txfonts}
\usepackage{mathptmx}
\usepackage[pdflang={en-US},pdftex]{hyperref}
\usepackage{color}
\usepackage{booktabs}
\usepackage{textcomp}

\usepackage{microtype}        
\usepackage{ccicons}          

\usepackage{todonotes}

\def\plaintitle{Improving Assessment on MOOCs Through Peer Identification and Aligned Incentives}
\def\plainauthor{Dilrukshi Gamage, Mark Whiting, Thejan Rajapakshe, Haritha Thilakarathne, Indika Perera, Shantha Fernando}
\def\plainkeywords{Massive Open Online Course; Peer Review}

\makeatletter
\def\url@leostyle{%
  \@ifundefined{selectfont}{
    \def\UrlFont{\sf}
  }{
    \def\UrlFont{\small\bf\ttfamily}
  }}
\makeatother
\def\pprw{8.5in}
\def\pprh{11in}

\definecolor{linkColor}{RGB}{6,125,233}
\hypersetup{%
  pdftitle={\plaintitle},
  pdfauthor={\plainauthor},
  pdfkeywords={\plainkeywords},
  pdfdisplaydoctitle=true, 
  bookmarksnumbered,
  pdfstartview={FitH},
  colorlinks,
  citecolor=black,
  filecolor=black,
  linkcolor=black,
  urlcolor=linkColor,
  breaklinks=true,
  hypertexnames=false
}

\begin{document}

\title{\plaintitle}

\numberofauthors{6}
\author{%
  \alignauthor{Dilrukshi Gamage\\
    \affaddr{University of Moratuwa}\\
    \affaddr{Katubedda, Sri Lanka}\\
    \email{dilrukshi.gamage@gmail.com}}\\
  \alignauthor{Mark E. Whiting\\
    \affaddr{Carnegie Mellon University}\\
    \affaddr{Pittsburgh, Pennsylvania}\\ 
    \email{mwhiting@andrew.cmu.edu}}\\
  \alignauthor{Thejan Rajapakshe\\
    \affaddr{Rajarata University of Sri Lanka}\\
    \affaddr{Anuradhapura, Sri Lanka}\\
    \email{coder.clix@gmail.com}}\\
  \alignauthor{Haritha Thilakarathne\\
    \affaddr{Rajarata University of Sri Lanka}\\
    \affaddr{Anuradhapura, Sri Lanka}\\
    \email{harithalht@gmail.com}}\\
  \alignauthor{Indika Perera\\
    \affaddr{University of Moratuwa}\\
    \affaddr{Katubedda, Sri Lanka}\\
    \email{indika@cse.mrt.ac.lk}}\\
  \alignauthor{Shantha Fernando\\
    \affaddr{University of Moratuwa}\\
    \affaddr{Katubedda, Sri Lanka}\\
    \email{shantha@cse.mrt.ac.lk}}\\
}

\maketitle

\begin{abstract}
Massive Open Online Courses (MOOCs) use peer assessment to grade open ended questions at scale, allowing students to provide feedback. Relative to teacher based grading, peer assessment on MOOCs traditionally delivers lower quality feedback and fewer learner interactions. We present the \textit{identified peer review} (IPR) framework, which provides non-blind peer assessment and incentives driving high quality feedback. We show that, compared to traditional peer assessment methods, IPR leads to significantly longer and more useful feedback as well as more discussion between peers. 
\end{abstract}

\category{K.3.1.}{COMPUTERS AND EDUCATION}{Computer Uses in Education}


\section{Introduction}
Peer assessment in Massive Open Online Courses (MOOCs) affords grading open ended assignments of many students, but this approach often can't provide the level of feedback that students need. Alternatives such as automated grading enable grading at scale, but require specialized assignment design to facilitate accurate algorithmic judgment, and can't deal well with open ended task designs, sacrificing student learning for ease of grading~\cite{kulkarni2015peer}. 

\begin{figure*}
 \centering
 \includegraphics[width=2\columnwidth]{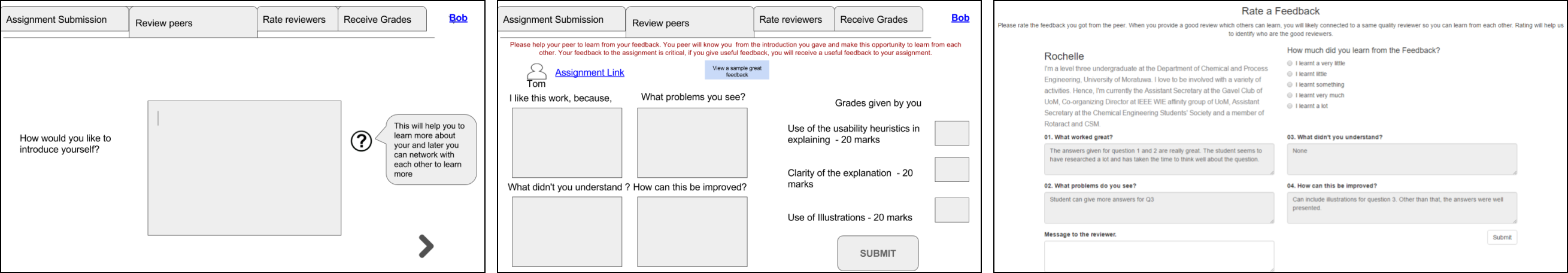}
 \caption{The interfaces of IPR: (1) the peer introduction form where reviewers provide a short introduction to be shared with those reading their reviews, (2) the review form where the four review prompts and grades are entered, (3) the feedback form displaying the feedback and prompting the receiver to rate its usefulness. Grades are not shown in this figure, but would appear after all feedback and ratings had been conducted.}
 \label{fig:iprInterface}
\end{figure*}

Peer assessment on MOOCs often leads to inaccurate grades and low quality feedback~\cite{reily2009two}, due to laziness, collusion, dishonesty, retaliation and a lack of time, experience or interest~\cite{kulkarni2015peer}. To counter this, humanizing feedback prompt phrasing~\cite{kotturi2015structure} and offering bonus points to feedback deemed helpful by the receiver~\cite{staubitz2016improving} have offered improved efficacy. In point systems, if a reviewer provided helpful feedback, the receivers will rate it, however without intrinsic motivation to provide a rating, this leads to single sides systems.

Additionally, MOOC peer reviews are carried out double blind where peers are not made aware of whom they are reviewing or who has reviewed them. This leads to a rising problem in MOOC assessment where due to increased anonymity, there is reduced accountability and eroded community affiliation~\cite{lu2007comparison}. Blind peer grading has been practiced in many face to face classroom environments and online learning environments to prevent grading bias and mitigate targeted criticism or bullying. A disadvantage of blind reviews arises in MOOCs when students provide lower quality and less insightful feedback because, being anonymous in review, they are not socially accountable. 

Evaluating peers work is a great means of learning. In face to face classroom situations peer evaluation often leads to a conversation where both parties interact richly and gain important understandings about the work, through back and forth communication. MOOCs provide forums for communication and networking, however they are often flooded with community discussion, weeding out paired peer discussions because individual social connections are not readily made. 

We introduce the \textit{Identified Peer Review} (IPR) framework with reduced anonymity and matched incentives to counter these issues of current approaches and increase communication. We evaluated the design with an between subject field experiment. This paper presents preliminary results with a small number of participants (n=87), and describes a future implementation and further evaluation plans.

\section{Related Work}
Assessment is important to the pedagogy of MOOCs
~\cite{bayne2013pedagogy}. MOOCs utilize peer assessment to assess student's work in a scalable way, not dependent on a one to many teacher to student relationship, but, optimal methods of choosing graders and assignments to grade remains an open question~\cite{balfour2013assessing}. The benefits of peer assessment include improvement of higher order thinking skills, consolidation of topical knowledge, and individualized feedback for each participant~\cite{gehringer2000strategies}. Giving and getting feedback has been identified as an effective way to learn in online~\cite{schank1999learning} and in the classroom. 

However, with the potential benefits, peer grading still faces challenges. Providing accurate grading where the performance of a novice is being judged by other novices is problematic. Students mistrust peer grades and anonymity lead students to be unscrupulous. Reviews are often short and not well considered, which is especially problematic when students work hard on an assignment and receive comments they can't learn from~\cite{suen2014peer}.

Our framework is design based on assessment interactions with visible incentivized peers. Anonymity is commonly practiced in online and in person class peer review systems. Due to the comfort of peers in providing critical feedback~\cite{bostock2000student}, the bigger problem in online reviewing is due to increased anonymity and reduced community affiliation~\cite{hamer2005method, lu2007comparison}. Visible identities lead to more constructive feedback than anonymous ones~\cite{vinther2012same}. Thus through providing non blind review, our framework may encourage accountable reviewing.

Feedback improves performance by changing students' locus of attention, focusing them on productive aspects of their work~\cite{kluger1996effects, kulkarni2015peer}. However, not all peers provide great feedback and some leave limited comments with no coherent message for improvement, or they are rogue reviews~\cite{reily2009two}. Rogue reviews are insufficient reviews caused by laziness, collusion, dishonesty, retaliation, competition, or malevolence~\cite{knight2011adapting}. To improve on this, PeerStudio~\cite{kulkarni2015peerstudio} peer assessment system is designed to encourage more feedback comments by showing short tips for writing comments just below the comment box. For example, if a response has no constructive feedback, it may remind students with phrases like: ''Quick check: Is your feedback actionable?'' by triggering heuristics word count on feedback~\cite{kulkarni2015peerstudio}. Students see such comments as more useful than rubrics in reviewing~\cite{kotturi2015structure}. Similar techniques are used to improve the quality of product reviews online~\cite{kim2006automatically}. Our framework uses a simple interface design reflecting these lessons by integrating four pointed questions with separate response areas.

In MOOCs, social connections and networking are attractive features for improving learning and providing a platform for desperate learners to interact and learn together. Since MOOCs are often open to the public, the diverse nature of students can be an asset for improving learning performance, increasing innovations, creativity and critical thinking rather than negativity~\cite{admiraal2014self, knight2011adapting}. The cMOOC~\cite{siemens2005connectivism} approach explains peers as a learning source and finds that the more connections achieved in network with diverse perspectives from participants leads to a richer learning environment. By providing feedback and interactions in viewing assignments, projects, and online discussions as opportunities for crowd-sourcing, this approach leads to superior results that otherwise cannot be achieved individually by students (or the instructor)~\cite{siemens2005connectivism}. However, the cMOOC approach requires significant integration and system familiarity, so it is not widespread. Peer assessment is a great way of learning from each others in the community, yet MOOC designs tend to be so complex that no student can simply choose a peer to align with for feedback. Our framework provides a connection and communication mechanism where peers can talk and learn directly. Such discussions and interactions with diverse groups increase the learning significantly~\cite{kulkarni2015talkabout}. In particular, an HCI course on Coursera encouraged students to post assignments to forums in getting feedback, leading to a conversation of feedback with identified peers and to more connectivity on fellow peers. In this HCI MOOC, more than 75\% students were in favor of sharing their assignments on the public~\cite{kulkarni2015peer}, where they assert that blind peer review only drives decreased social connection and availability of diverse perspectives.

\section{Identified Peer Reviewing Framework}
The IPR framework is built upon 4 phases: submit assignments, review peers, rate feedback, receive grades (Figure~\ref{fig:iprInterface}). In the first phase, students submit their assignments. In the second phase, students are given a \textit{peer introduction form} to provide a short public introduction to be displayed in reviews they conduct. Next students review peer's work with the \textit{review form}, an interface with four targeted feedback fields and grade input field. Students then receive feedback on their own work from other peers in the \textit{feedback form}, which initially shows only the feedback fields, a text input to converse with the reviewer, and a five point scale to rate the usefulness of the feedback. Once all rating is finished, students will receive the grades their reviewers assigned them. 

Each student reviews three other students' work, and receives reviews from three students. Unlike most peer reviewing systems, IPR does not randomize peer review assignment, instead aligning incentives by matching a student with reviewers based on the usefulness of that student's previous feedback. Students are motivated to provide their peers with high quality feedback to get high quality peer feedback in return. The next round of review allocations are made when feedback has been rated. Only after a student has received and rated all their feedback are grades made visible, so good feedback ratings can't be bought with inflated grades.

\section{Evaluation and Results}
We hypothesize that having identifiable peers and matched incentives will increase the feedback quality and lead to more communication than blind peer reviewing. To test this, we conducted a between subject experiment with 3 conditions.

 \begin{enumerate}
 \item \textbf{Control condition}: Blind peer review, randomized reviewers.
 \item \textbf{IPR random condition}: Identified peer review, randomized reviewers.
 \item \textbf{IPR incentive condition}: Identified peer review, incentivized reviewers. 
 \end{enumerate}

87 participants collected from an online advertisement placed through authors social media channels and participants were randomly assigned one of three conditions. We created an assignment on the subject ''Creativity and Innovation'', requiring no previous lessons or special subject knowledge. We measured how students perform in each condition, and how useful the feedback was to them.

\subsection{How does reviewer blindness influence feedback quality?}
To test this, we ran an independent t-test using the control condition and the identified condition. Students in the control condition had an average feedback quality score of $1.21$ out of $29$ ($\sigma= .243$), and students in the identified condition had an average score of $2.43$ out of $29$ ($\sigma=1.02$). An unpaired t-test confirmed that students in the identified condition is significant than other: $t(56)= 6.24$, $p = 0.000$, $\alpha=0.05$. Which means identified peer reviewing had a large positive effect on the usefulness of feedback compared with double-blind peer-reviewing of assignments. 

\subsection{How does incentive alignment influence feedback quality?}
To test this we ran another independent t-test using the identified condition and the aligned incentives condition. Students in the identified condition had an average feedback quality score of $2.43$ out of $29$ ($\sigma = 1.08$), and students in the aligned incentives condition had an average score of $2.12$ out of $25$ ($\sigma = 0.71$). The t-test confirmed that students in the identified peer grading condition is significant: $t(48) = 1.19$, $p = 0.009$, $\alpha = 0.05$.

A possible impact of the incentive alignment condition would be to make the gap between the best and worst students larger as the best students help each other, and the worst students are never given high quality feedback. In this initial study, we have not seen evidence of this effect, and hope to explore this in more detail in the future.

\subsection{Students motivation to communicate}
We analyzed the peer reviews done by all 3 conditions from the communicating message box shown in Figure~\ref{fig:iprInterface}. The control group blind peer reviews contained only 29 messages responding to the feedback they received while the identified condition contained 53 and the aligned incentives condition received 49. The control condition group messages did not contain meaningful communication while treatment groups both had students interested in further communicating together. For example control case messages were limited to one or two words such as ''Thanks, OK'' while the treatment groups had messages like ''I like to network too, it is great to be connected to someone new, thanks for the advice - KIT.'' The message box also gave students means to ask and respond to questions about the assignment or feedback, for example, a student was able to inform their reviewer that a broken link to the assignment for review had been repaired. 

\section{Discussion and Future work}
We aimed to reintroduce identity in a humanize form where peers introduce themselves to each other. This shared identity showed value, driving conversation and improving feedback quality. Inspired by the design studio concept ~\cite{tinapple2013critviz} and cMOOCs where learners make meaningful connections and learn by giving feedback, we designed the IPR framework and witness early results of students' response, compared to a control condition. At the same time adding incentive matching in which reviewers and those being reviewed both had the incentive to be honest in their responses. We believe these results introduce incentive compatible interaction design~\cite{gaikwad2016boomerang} to MOOCs and that this kind of design offers significant opportunities for ongoing improvement in this field.

This research will continue through testing the influence IPR has on larger cohorts' in a fully develop MOOC. For decades, online peer reviewing has been blind, we aimed to integrate identity in assessment on an MOOC, by carefully designing interactions leading students to discuss and have reason to improve their feedback quality. Initial experimentation showed minimal advantage using incentive alignment over randomized review assignment, but both test conditions performed significantly better in review usefulness and in starting conversations. We believe this is a consequence of the small sample size, so we will explore efficiency with a larger cohort where our ultimate goal is to provide effective feedback to students with meaningful connections, so they can benefit from the diverse crowd in a MOOC.

\balance{}
\bibliographystyle{SIGCHI-Reference-Format}
\bibliography{refs}


\begin{thebibliography}{00}


\ifx \showCODEN    \undefined \def \showCODEN     #1{\unskip}     \fi
\ifx \showDOI      \undefined \def \showDOI       #1{{\tt DOI:}\penalty0{#1}\ }
  \fi
\ifx \showISBNx    \undefined \def \showISBNx     #1{\unskip}     \fi
\ifx \showISBNxiii \undefined \def \showISBNxiii  #1{\unskip}     \fi
\ifx \showISSN     \undefined \def \showISSN      #1{\unskip}     \fi
\ifx \showLCCN     \undefined \def \showLCCN      #1{\unskip}     \fi
\ifx \shownote     \undefined \def \shownote      #1{#1}          \fi
\ifx \showarticletitle \undefined \def \showarticletitle #1{#1}   \fi
\ifx \showURL      \undefined \def \showURL       #1{#1}          \fi

\bibitem{admiraal2014self}
{Wilfried Admiraal}, {Bart Huisman}, {and} {Maarten Van~de Ven}. 2014.
\newblock \showarticletitle{Self-and peer assessment in massive open online
  courses}.
\newblock {\em International Journal of Higher Education\/} {3}, 3 (2014),
  p119.
\newblock


\bibitem{balfour2013assessing}
{Stephen~P Balfour}. 2013.
\newblock \showarticletitle{Assessing writing in MOOCs: Automated essay scoring
  and calibrated peer review (tm)}.
\newblock {\em Research \& Practice in Assessment\/}  {8} (2013).
\newblock


\bibitem{bayne2013pedagogy}
{Sian Bayne} {and} {Jen Ross}. 2013.
\newblock The Pedagogy of the Massive Open Online Course (MOOC): the UK View,
  the Higher Education Academy.
\newblock   (2013).
\newblock


\bibitem{bostock2000student}
{Stephen Bostock}. 2000.
\newblock \showarticletitle{Student peer assessment}.
\newblock {\em Learning Technology\/} (2000).
\newblock


\bibitem{gaikwad2016boomerang}
{Snehalkumar Neil~S Gaikwad}, {Durim Morina}, {Adam Ginzberg}, {Catherine
  Mullings}, {Shirish Goyal}, {Dilrukshi Gamage}, {Christopher Diemert},
  {Mathias Burton}, {Sharon Zhou}, {Mark Whiting}, {and} {others}. 2016.
\newblock \showarticletitle{Boomerang: Rebounding the Consequences of
  Reputation Feedback on Crowdsourcing Platforms}. In {\em Proceedings of the
  29th Annual Symposium on User Interface Software and Technology}. ACM,
  625--637.
\newblock


\bibitem{gehringer2000strategies}
{Edward~F Gehringer}. 2000.
\newblock \showarticletitle{Strategies and mechanisms for electronic peer
  review}. In {\em Frontiers in Education Conference, 2000. FIE 2000. 30th
  Annual}, Vol.~1. IEEE, F1B--2.
\newblock


\bibitem{hamer2005method}
{John Hamer}, {Kenneth~TK Ma}, {and} {Hugh~HF Kwong}. 2005.
\newblock \showarticletitle{A method of automatic grade calibration in peer
  assessment}. In {\em Proceedings of the 7th Australasian conference on
  Computing education-Volume 42}. Australian Computer Society, Inc., 67--72.
\newblock


\bibitem{kim2006automatically}
{Soo-Min Kim}, {Patrick Pantel}, {Tim Chklovski}, {and} {Marco Pennacchiotti}.
  2006.
\newblock \showarticletitle{Automatically assessing review helpfulness}. In
  {\em Proceedings of the 2006 Conference on empirical methods in natural
  language processing}. Association for Computational Linguistics, 423--430.
\newblock


\bibitem{kluger1996effects}
{Avraham~N Kluger} {and} {Angelo DeNisi}. 1996.
\newblock \showarticletitle{The effects of feedback interventions on
  performance: a historical review, a meta-analysis, and a preliminary feedback
  intervention theory.}
\newblock {\em Psychological bulletin\/} {119}, 2 (1996), 254.
\newblock


\bibitem{knight2011adapting}
{Linda~V Knight} {and} {Theresa~A Steinbach}. 2011.
\newblock \showarticletitle{Adapting peer review to an online course: An
  exploratory case study}.
\newblock {\em Journal of Information Technology Education\/}  {10} (2011),
  81--100.
\newblock


\bibitem{kotturi2015structure}
{Yasmine Kotturi}, {Chinmay~E Kulkarni}, {Michael~S Bernstein}, {and} {Scott
  Klemmer}. 2015.
\newblock \showarticletitle{Structure and messaging techniques for online peer
  learning systems that increase stickiness}. In {\em Proceedings of the Second
  (2015) ACM Conference on Learning@ Scale}. ACM, 31--38.
\newblock


\bibitem{kulkarni2015talkabout}
{Chinmay Kulkarni}, {Julia Cambre}, {Yasmine Kotturi}, {Michael~S Bernstein},
  {and} {Scott~R Klemmer}. 2015b.
\newblock \showarticletitle{Talkabout: Making distance matter with small groups
  in massive classes}. In {\em Proceedings of the 18th ACM Conference on
  Computer Supported Cooperative Work \& Social Computing}. ACM, 1116--1128.
\newblock


\bibitem{kulkarni2015peer}
{Chinmay Kulkarni}, {Koh~Pang Wei}, {Huy Le}, {Daniel Chia}, {Kathryn
  Papadopoulos}, {Justin Cheng}, {Daphne Koller}, {and} {Scott~R Klemmer}.
  2015c.
\newblock \showarticletitle{Peer and self assessment in massive online
  classes}.
\newblock In {\em Design thinking research}. Springer, 131--168.
\newblock


\bibitem{kulkarni2015peerstudio}
{Chinmay~E Kulkarni}, {Michael~S Bernstein}, {and} {Scott~R Klemmer}. 2015a.
\newblock \showarticletitle{PeerStudio: rapid peer feedback emphasizes revision
  and improves performance}. In {\em Proceedings of the Second (2015) ACM
  Conference on Learning@ Scale}. ACM, 75--84.
\newblock


\bibitem{lu2007comparison}
{Ruiling Lu} {and} {Linda Bol}. 2007.
\newblock \showarticletitle{A comparison of anonymous versus identifiable
  e-peer review on college student writing performance and the extent of
  critical feedback}.
\newblock {\em Journal of Interactive Online Learning\/} {6}, 2 (2007),
  100--115.
\newblock


\bibitem{reily2009two}
{Ken Reily}, {Pam~Ludford Finnerty}, {and} {Loren Terveen}. 2009.
\newblock \showarticletitle{Two peers are better than one: aggregating peer
  reviews for computing assignments is surprisingly accurate}. In {\em
  Proceedings of the ACM 2009 international conference on Supporting group
  work}. ACM, 115--124.
\newblock


\bibitem{schank1999learning}
{Roger~C Schank}, {Tamara~R Berman}, {and} {Kimberli~A Macpherson}. 1999.
\newblock \showarticletitle{Learning by doing}.
\newblock {\em Instructional-design theories and models: A new paradigm of
  instructional theory\/}  {2} (1999), 161--181.
\newblock


\bibitem{siemens2005connectivism}
{George Siemens}. 2005.
\newblock \showarticletitle{Connectivism: Learning as network-creation}.
\newblock {\em ASTD Learning News\/} {10}, 1 (2005).
\newblock


\bibitem{staubitz2016improving}
{Thomas Staubitz}, {Dominic Petrick}, {Matthias Bauer}, {Jan Renz}, {and}
  {Christoph Meinel}. 2016.
\newblock \showarticletitle{Improving the Peer Assessment Experience on MOOC
  Platforms}. In {\em Proceedings of the Third (2016) ACM Conference on
  Learning@ Scale}. ACM, 389--398.
\newblock


\bibitem{suen2014peer}
{Hoi~K Suen}. 2014.
\newblock \showarticletitle{Peer assessment for massive open online courses
  (MOOCs)}.
\newblock {\em The International Review of Research in Open and Distributed
  Learning\/} {15}, 3 (2014).
\newblock


\bibitem{tinapple2013critviz}
{David Tinapple}, {Loren Olson}, {and} {John Sadauskas}. 2013.
\newblock \showarticletitle{CritViz: Web-based software supporting peer
  critique in large creative classrooms}.
\newblock {\em Bulletin of the IEEE Technical Committee on Learning
  Technology\/} {15}, 1 (2013), 29.
\newblock


\bibitem{vinther2012same}
{Siri Vinther}, {O Haagen~Nielsen}, {Jacob Rosenberg}, {Niels Keiding}, {and}
  {TV Shroeder}. 2012.
\newblock \showarticletitle{Same review quality in open versus blinded peer
  review in" Ugeskrift for L{\ae}ger}.
\newblock {\em Dan Med J\/} {59}, 8 (2012), A4479.
\newblock


\end{thebibliography}

\end{document}